\documentclass[%
 aip,
 amsmath,amssymb,
 reprint,%
]{revtex4-1}

\usepackage{graphicx}
\usepackage{dcolumn}
\usepackage{bm}

\usepackage{cancel}
\usepackage[monochrome]{xcolor}

\usepackage[utf8]{inputenc}
\usepackage[T1]{fontenc}
\usepackage{mathptmx}
\usepackage{etoolbox}

\newcommand{\RomanNumeralCaps}[1]{\MakeUppercase{\romannumeral #1}}

\makeatletter
\def\@email#1#2{%
 \endgroup
 \patchcmd{\titleblock@produce}
  {\frontmatter@RRAPformat}
  {\frontmatter@RRAPformat{\produce@RRAP{*#1\href{mailto:#2}{#2}}}\frontmatter@RRAPformat}
  {}{}
}%
\makeatother
\begin{document}

\title{Kinetic pathways of coesite densification from metadynamics}

\author{David Vrba}
\author{Roman Martoňák}
\thanks{The authors to whom correspondence may be addressed: david.vrba@fmph.uniba.sk, martonak@fmph.uniba.sk}
\affiliation{Department of Experimental Physics, Faculty of Mathematics, Physics and Informatics, Comenius University, Mlynsk\'{a} Dolina F2, 842 48 Bratislava, Slovakia}

\date{\today}

\begin{abstract}
We study compression of coesite to pressures above 35 GPa, substantially beyond the equilibrium transition pressure to octahedral phases (8 GPa to stishovite).
Experiments at room temperature showed that up to 30 GPa the metastable coesite structure develops only minor displacive changes (coesite-\RomanNumeralCaps{2} and coesite-\RomanNumeralCaps{3}) while the Si atoms remain 4-coordinated. Beyond 30 GPa, reconstructive transformations start, following different pathways from the complex structure of coesite. Besides amorphization, two different crystalline outcomes were observed. One is formation of defective high-pressure octahedral phases (Hu et al., 2015) and another one is formation of
unusual and complex dense phases coesite-\RomanNumeralCaps{4} and coesite-\RomanNumeralCaps{5} with Si atoms in 4-fold, 5-fold and 6-fold coordination (Bykova et al., 2018). 
Capturing these structural transformations computationally represents a challenge. Here we show that employing metadynamics with Si-O coordination number and volume as generic collective variables in combination with a machine-learning based ACE potential (Erhard et al., 2024), one naturally observes all three mentioned pathways, resulting in the phases observed experimentally. We describe the atomistic mechanisms along the transformation pathways. While the pathway to coesite-\RomanNumeralCaps{4} is simpler, the transformation to octahedral phases involves two steps: first, a hcp sublattice of O atoms is formed where Si atoms occupy octahedral positions but the octahedra chains do not form a regular pattern. In the second step, the Si atoms order and the chains develop a more regular arrangement.
We predict that the pathway to coesite-\RomanNumeralCaps{4} is preferred at room temperature, while at 600 K the formation of octahedral phases is more likely. 
\end{abstract}

\maketitle

\section{Introduction}

Silica (SiO$_2$) in its crystalline as well as amorphous forms represents a unique material for a number of reasons. Due to its abundance in the Earth's crust and presence also in other planets, it is highly important for geophysics as well as planetary physics. At the same time it is a highly important material for industrial use (glass, microelectronics, etc.). 
From a more fundamental point of view, its rich polymorphism and existence of the archetypal amorphous phase make silica a prominent example of a system with numerous structural transitions which can be induced by pressure and temperature\cite{dubrovinsky2004}. The structural transformations in a covalent network can be naturally classified into displacive and reconstructive. The temperature-induced transitions typically involve small displacive changes of the relatively rigid constituent coordination polyhedra (typically rotations). On the other hand, the pressure-induced transitions are often reconstructive, including a change of coordination, allowing the densification of the system due to reconstruction of the bonding network. 

The pressure-induced reconstructive transitions are thermodynamically typically first order, and the involved phases are separated by substantial free-energy barriers. For this reason, one often observes strong kinetic effects and hysteresis, allowing the phases to exist in a metastable form beyond the equilibrium phase boundary, especially in experiments conducted at room or lower temperatures. At the same time, metastable and disordered phases are often created as a result of the structural transformations, pointing to the difficulties in rearranging the atoms to the optimal configuration. A well-known example is the compression of $\alpha$-quartz, which should under equilibrium conditions transform to coesite at 2 GPa. However, this transition is very slow, even at higher temperatures such as $800$ K, lasting for hours \cite{Bohlen1982}. At room temperature $\alpha$-quartz remains metastable up to $21$ GPa, where it transforms into a metastable phase with mixed coordination known as quartz II \cite{Kingma1993b}, consisting of tetrahedra and octahedra \cite{Choudhury2006}. 

Coesite represents the equilibrium phase of silica in the pressure region $2-8$ GPa and features a substantial densification of $12$ \% with respect to $\alpha$-quartz, while retaining the 4-coordinated Si atoms, arranged in corner-sharing tetrahedra. This is achieved via a complicated reconstruction of the covalent network resulting in a large unit cell with $48$ atoms ($Z=16$), space group $C2/c$. The complexity is evident also in the existence of several kinds of non-equivalent atoms ($2$ kinds for Si and $5$ kinds for O atoms). On the other hand, the next equilibrium structure, stishovite, features a much simpler tetragonal structure ($Z=2$), space group $P4_{2}/mnm$, consisting of straight chains of edge-sharing octahedra \cite{Teter1998}. Taking into account the pronounced structural difference between coesite and stishovite, it is natural to expect also in this case both strong hysteresis and metastability.

Coesite compression was first studied experimentally in Ref.\cite{Hemley1988} and found to amorphize at $34$ GPa. Later in Ref.\cite{Kingma1993a}, it was argued that the observed amorphous phase was actually not fully disordered, retaining remnants of original crystalline order. Over the last two decades, compression of coesite was studied both experimentally and computationally. 
On the computational side, compression of coesite was studied by ab initio metadynamics\cite{Laio2002} scheme for simulations of structural phases transitions, employing the supercell vectors as collective variables (CVs)\cite{Martonak2006, Martonak2007}. Using a single unit cell with 48 atoms, coesite transformed at $22$ GPa and $600$ K to perfect post-stishovite seifertite ($\alpha$-PbO$_2$-like SiO$_2$). 
In Refs.\cite{Cernok2014a, Cernok2014b} the compression of coesite was studied by in situ Raman spectroscopy or single-crystal X-ray diffraction at room temperature and pressures up to $50$ GPa. 
Two displacive phase transitions of the metastable coesite structure were found.
The first phase transition occurs around $23$ GPa, resulting in coesite-\RomanNumeralCaps{2}.
This is achieved via a rearrangement of tetrahedra, creating two different $4$-membered rings, which results in modulations along the \textit{y} direction, with a larger unit cell containing $96$ atoms ($Z=32$), space group $P2_1/n$.
The second phase transition occurs around $31$ GPa, transforming to coesite-\RomanNumeralCaps{3}. 
The unit cell of coesite-\RomanNumeralCaps{3} contains $72$ atoms ($Z=24$), space group $P\text{-}1$, with modulations along the \textit{y} direction.
Both transformations are reversible upon decompression, and coesite-\RomanNumeralCaps{2} and coesite-\RomanNumeralCaps{3} revert to coesite-\RomanNumeralCaps{1} at ambient pressure.
In Ref.\cite{Wu2018}, compression of coesite was studied by X-ray diffraction and upon compression above $36$ GPa
a different phase transition from coesite-\RomanNumeralCaps{2} was observed, which did not result in coesite-\RomanNumeralCaps{3}.
The structure with commensurate modulation along the \emph{y} axis was denoted as coesite-\RomanNumeralCaps{10}.
However, due to the sample's low crystallinity, the structure could not be determined.

In Ref.\cite{Hu2015}, 
a combined experimental and \textit{ab initio} metadynamics study was presented. Using a single-crystal synchrotron X-ray diffraction, it was shown that coesite upon compression to 40 GPa transformed into a dense crystalline phase. Based on \textit{ab initio} metadynamics simulation using the Si-O coordination number and two lattice parameters as CVs, a monoclinic octahedral phase (\textit{P2/c}) was found which could explain the basic features of the observed diffraction pattern. The computational study\cite{Liu2017} performed molecular dynamics simulations of coesite compression with a classical potential to pressures beyond 50 GPa and identified besides amorphization also a formation of a high-pressure octahedral phase (HPO) which represented a defective $\alpha$-PbO$_2$-like SiO$_2$ with 2$\times$2 zigzag chains. The XRD pattern of this phase was similar to the one found experimentally in Ref.\cite{Hu2015}.

In Ref.\cite{Bykova2018} the compression of coesite was studied by single-crystal X-ray diffraction at room temperature and pressures up to 70 GPa. Upon compression up to 30 GPa coesite-\RomanNumeralCaps{2} and coesite-\RomanNumeralCaps{3} was formed.
Beyond 30 GPa reconstructive transformations take place, creating highly complex dense phases coesite-\RomanNumeralCaps{4} and coesite-\RomanNumeralCaps{5} (both $Z = 16$ and space group $P\bar{1}$) with Si atoms in 4-fold, 5-fold and 6-fold coordination, featuring non-standard connections of octahedra violating the 3$^{rd}$ Pauling rule. These structures were found to have unusually high enthalpies, about 0.39 eV/atom above stishovite at 38 GPa, reflecting the persistent existence of low-coordinated Si atoms at conditions of strong overpressurization. 

Capturing these complex structural transformations computationally, however, is a non-trivial task. Clearly, plain MD is not able to cross the free-energy barriers and one needs an enhanced sampling method, such as, e.g., metadynamics\cite{Laio2002} (for review articles see, e.g., Refs.\cite{Barducci2011, Bussi2020}). Ideally, one should avoid making prior assumptions about the nature of the final phase, allowing for all possible outcomes. In this context, the proper and generic choice of the collective variables (CVs) is critically important. In Refs.\cite{Martonak2006,Martonak2007} the supercell vector components were used as CVs, describing the global changes of the crystal periodicity which indirectly reflect the structural changes. A better alternative is to focus directly on the local structural changes induced in the coordination spheres of individual atoms, which necessarily accompany the densification of the system.
In metadynamics performed in Ref.\cite{Hu2015} a combination of the Si-O coordination number and selected lattice parameters was used. The combination of coordination number and system volume was shown to work well for the B1/B2 transformation in NaCl\cite{Badin2021}. 
In the present work, we employ metadynamics with Si-O and Si-Si coordination numbers as well as volume as generic collective variables.  We show that this approach naturally finds both experimentally observed crystalline phases coesite-\RomanNumeralCaps{4} and coesite-\RomanNumeralCaps{5}, as well as the transformation to HPO. The paper is organized as follows. In section \ref{sec:meth} we introduce the method and its implementation.
The next section \ref{sec:res} is central and presents the results as well as their discussion. In the final section \ref{sec:conc} we draw conclusions. 

\section{Method}
\label{sec:meth}

For all MD simulations, we employed LAMMPS \cite{LAMMPS}. 
For enhanced sampling, we performed metadynamics simulations using the PLUMED package\cite{PLUMED}.
In MD simulations, we used a timestep $dt=0.002$ ps and fix npt.
The temperature was controlled using a \textcolor{blue}{Nose-Hoover thermostat \cite{Nose1984, Hoover1985}} with a damping parameter 
of $0.2$ ps. The pressure was controlled via a fully flexible cell using a triclinic  \textcolor{blue}{Parrinello-Rahman barostat \cite{Parrinello1981}}  with a damping parameter of $2.0$ ps.
This ensures that all volume as well as shape fluctuations of the supercell are allowed within the NPT ensemble.

As CVs, we mainly used the mean coordination number (CN) between Si and O atoms (the case of the Si-Si coordination related CVs is described in Supp.~Mat.) as well as the volume per SiO$_2$ unit. The coordination number and volume were employed using the PLUMED keywords \texttt{COORDINATION} and \texttt{VOLUME}, respectively. 
More details about the metadynamics method as well as values of specific parameters, related to switching functions, Gaussians and their deposition, etc., can be found in the Supp. Mat.

\textcolor{blue}{For the description of the system, we employ the recently developed ACE potential\cite{Erhard2024}, based on the Atomic Cluster Expansion formalism\cite{ACE_Drautz2019}. This potential was developed via the active-learning workflow (details of the database and training procedure can be found in Ref.\cite{Erhard2024}, Supp.~Mat.) and was shown to be accurate. Its high computational efficiency (GPU-capable) allows the simulation of much larger systems compared to \textit{ab initio} simulations.}

\textit{Ab initio} calculations were carried out using the Vienna Ab initio Simulation Package (VASP) \cite{Kresse1993, Kresse1996a, Kresse1996b, Kresse1999}. 
The projector augmented-wave (PAW) method \cite{Bloch1994} was employed, and the exchange-correlation energy was treated using the Perdew–Burke–Ernzerhof (PBE) functional\cite{Perdew1996}. We used the standard PAW pseudopotentials with $4$ valence electrons for Si (valence electron configuration $3$s$^2$ $3$p$^2$) and with $6$ valence electrons for O (valence electron configuration $2$s$^2$ $2$p$^4$).
Values of specific parameters for \textit{ab initio} calculations are listed in the Supp.~Mat. 

For visualization in Figs. \ref{fig:HPO_transition}, \ref{fig:HPO_chains} and \ref{fig:coesite4_transition} and analysis of atomic configurations in Figs. \ref{fig:hpo-cv} and \ref{fig:coesite4} we used OVITO\cite{OVITO}. \textcolor{blue}{The XRD patterns in Fig.6 were calculated using the VESTA package\cite{VESTA}.}

\section{Results and Discussion}
\label{sec:res}

\begin{figure}
\includegraphics[width=0.5\textwidth]{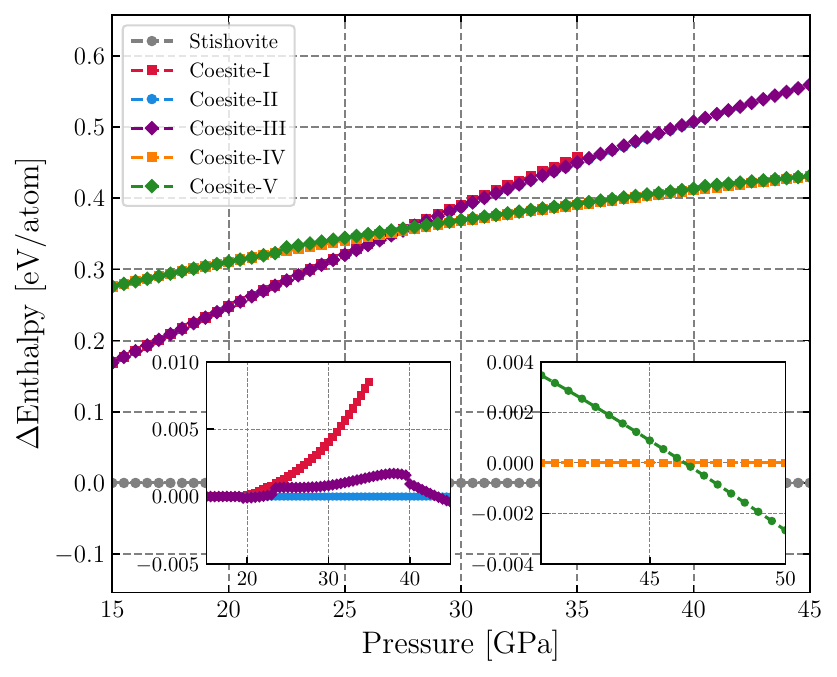}
\caption{\label{fig:enthalpy-ml-all} Difference of enthalpies between coesite-\RomanNumeralCaps{1}, \RomanNumeralCaps{2}, \RomanNumeralCaps{3}, \RomanNumeralCaps{4} and \RomanNumeralCaps{5} compared to stishovite. Enthalpy curves of coesite-\RomanNumeralCaps{1}, \RomanNumeralCaps{2} and \RomanNumeralCaps{3} cross the ones of \RomanNumeralCaps{4} and \RomanNumeralCaps{5} around $27$ GPa, while enthalpy curves of coesite-\RomanNumeralCaps{4} and \RomanNumeralCaps{5} cross around $46$ GPa. Inset on the left shows difference of enthalpies between coesite-\RomanNumeralCaps{1} and \RomanNumeralCaps{3} compared to coesite-\RomanNumeralCaps{2}. Inset on the right shows difference of enthalpies between coesite-\RomanNumeralCaps{4} and coesite-\RomanNumeralCaps{5}.}
\end{figure}

According to Ref.\cite{Bykova2018}, a suitable starting point to study reconstructive transitions from coesite to octahedral phases should be coesite-\RomanNumeralCaps{3}. However, the difference between coesite-\RomanNumeralCaps{2} and coesite-\RomanNumeralCaps{3} is in the present context of minor importance. It turned out that the ACE potential\cite{Erhard2024} we use actually does not properly reproduce the order of the modulated phases in the interval $20$ to $40$ GPa, as can be seen in Fig.\ref{fig:enthalpy-ml-all} (inset)\footnote{We note here that with the earlier version of the ML-QUIP potential$^{36}$ the order of phases coesite-\RomanNumeralCaps{2} and coesite-\RomanNumeralCaps{3} is correct, the transition pressure being $30$ GPa, and we were able to find coesite-\RomanNumeralCaps{3} by plain MD simulation at $35$ GPa and $300$ K.}.
Since the enthalpy of coesite-\RomanNumeralCaps{2} is marginally lower than that of coesite-\RomanNumeralCaps{3} in the pressure range from $23$ GPa to $40$ GPa, 
we chose coesite-\RomanNumeralCaps{2} as the starting structure for our metadynamics simulations, since it represents the more stable structure. 
We note that the enthalpy differences between coesite-\RomanNumeralCaps{1}, \RomanNumeralCaps{2}, and \RomanNumeralCaps{3} are of the order of a few meV per atom and these phases can be viewed as minor displacive variants of the parent phase coesite-\RomanNumeralCaps{1}. 
On the other hand, the phase transition from coesite-\RomanNumeralCaps{3} to coesite-\RomanNumeralCaps{4} involves significantly larger enthalpy differences, on the order of tens of meV per atom, reflecting bond breaking and reconstruction.
This difference in energy scales justifies the use of coesite-\RomanNumeralCaps{2} as a suitable starting point for exploring the reconstructive phase transitions at higher pressures.

We performed the metadynamics simulations at pressure of 35 GPa and temperatures of 300 K and 600 K. For each temperature we repeated the simulation 10 times under identical conditions, starting from the same initial structure of coesite-\RomanNumeralCaps{2} (supercell 4$\times$1$\times$4 with 1536 atoms, see Supp. Mat.) with different initialization of velocities. 
In simulations at 300 K, in one run we observed formation of an HPO structure, in three runs formation of coesite-\RomanNumeralCaps{4} and coesite-\RomanNumeralCaps{5}, and in six runs formation of a mixed or disordered structure. 
In simulations at 600 K, in one run we observed formation of coesite-\RomanNumeralCaps{4} and coesite-\RomanNumeralCaps{5}, in four runs formation of an HPO-like structure with varying order of chains (see later) 
and in five runs formation of a disordered structure. In the following we focus on the analysis of the runs at 300 K which resulted in pure crystalline phases.
The other outcomes represented either highly disordered or mixed structures, where part of the supercell was significantly more ordered than another (some of them are shown in Supp. Mat.). 
We now describe and analyze in detail the two important pathways, to HPO and to coesite-\RomanNumeralCaps{4}.

\subsection{Pathway to HPO}

\begin{figure}
\includegraphics[width=8cm]{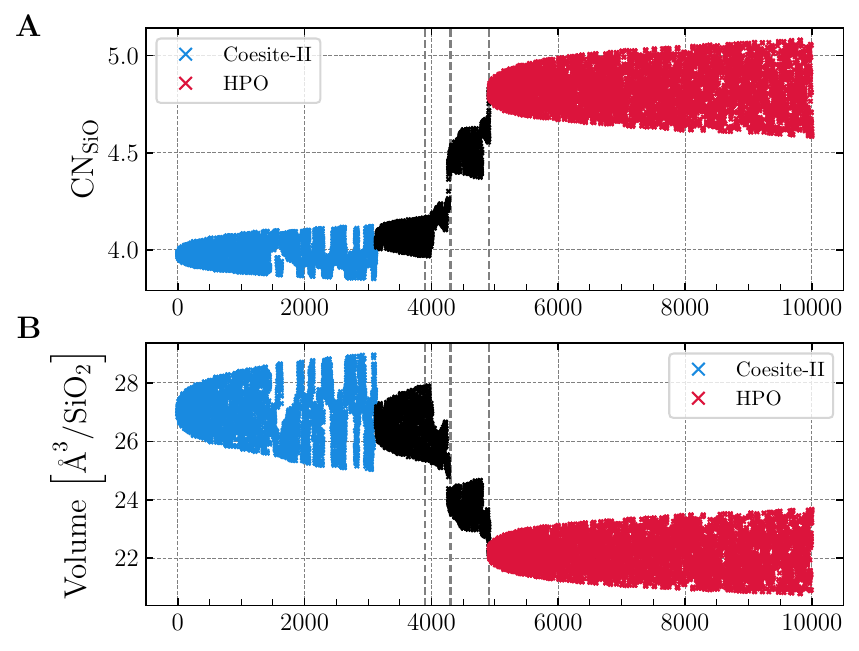}
\includegraphics[width=8cm]{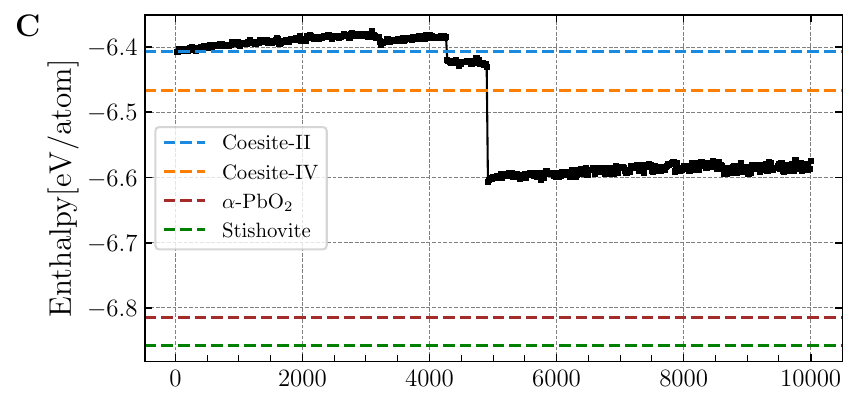}
\includegraphics[width=8cm]{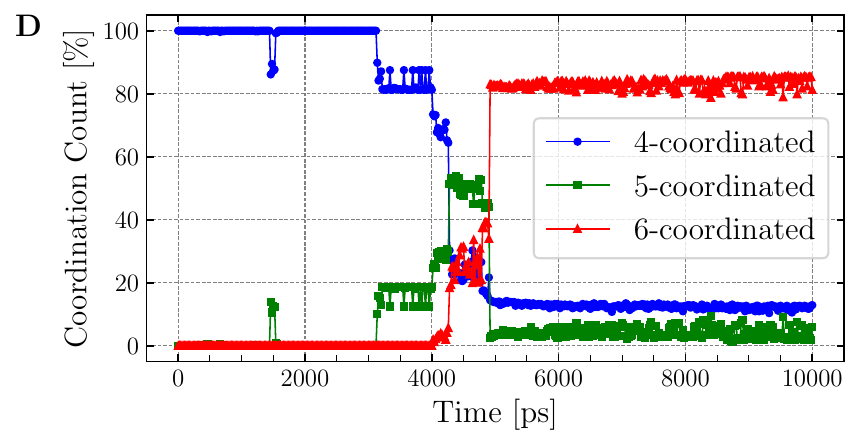}
\caption{\label{fig:hpo-cv} Evolution of the mean coordination number (\textbf{A}), volume (\textbf{B}), enthalpy (\textbf{C}) and number of 4-, 5- and 6-coordinated Si atoms (\textbf{D}) during the metadynamics simulation at $p=35$ GPa and $T=300$ K which resulted in formation of the HPO phase. In the evolution of CVs, the initial coesite-\RomanNumeralCaps{2} phase is shown in blue, the final HPO phase in red and intermediate structures in black.}
\end{figure}

\begin{figure}
\includegraphics[width=8cm]{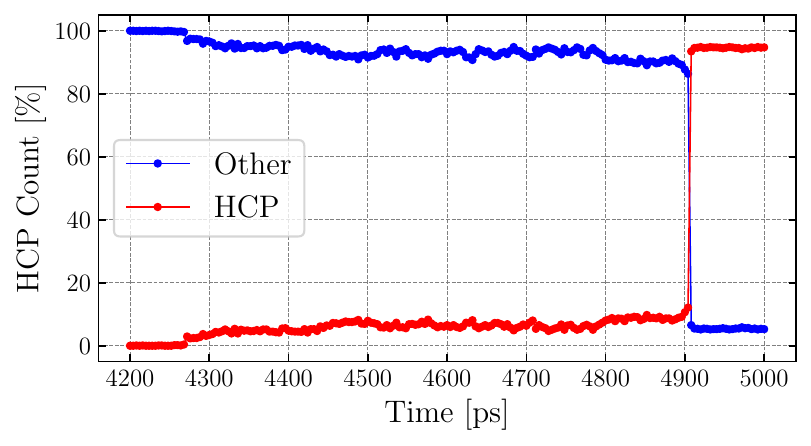}
\caption{\label{fig:hcp-count} Evolution of the fraction of oxygen atoms in a hcp lattice during formation of HPO phase at $p=35$ GPa and $T=300$ K. }
\end{figure}

\begin{figure*}
\includegraphics[width=\textwidth]{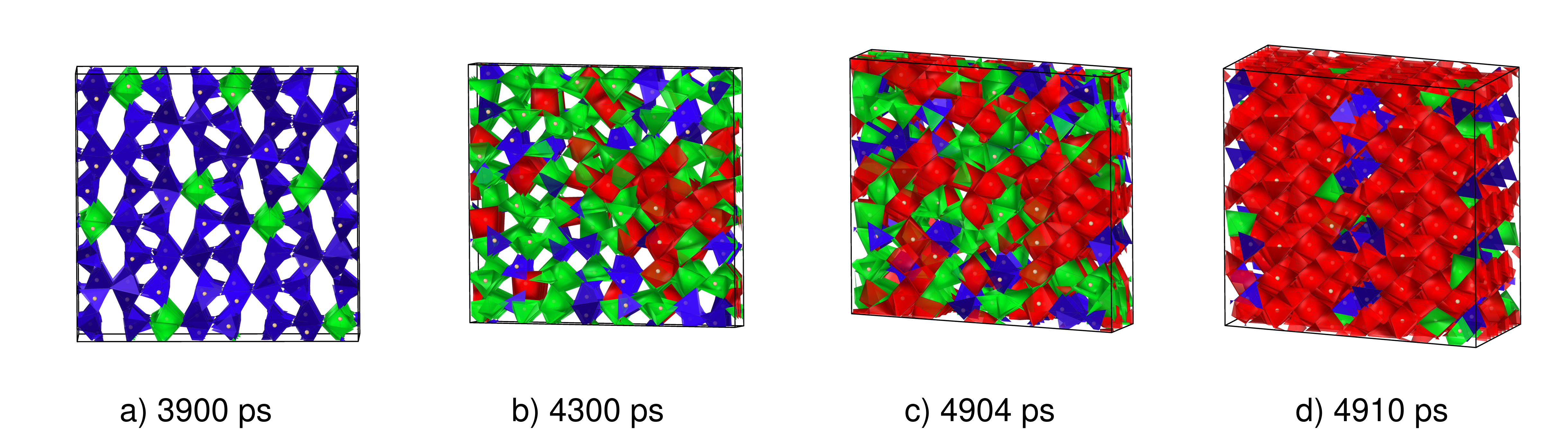}
\caption{\label{fig:HPO_transition} Evolution of the structure of the system during formation of the HPO phase at $p=35$ GPa and $T=300$ K. Si atoms with coordination of 4,5 and 6 are shown as blue, green and red polyhedra, respectively.}
\end{figure*}

\begin{figure}
\includegraphics[width=8cm]{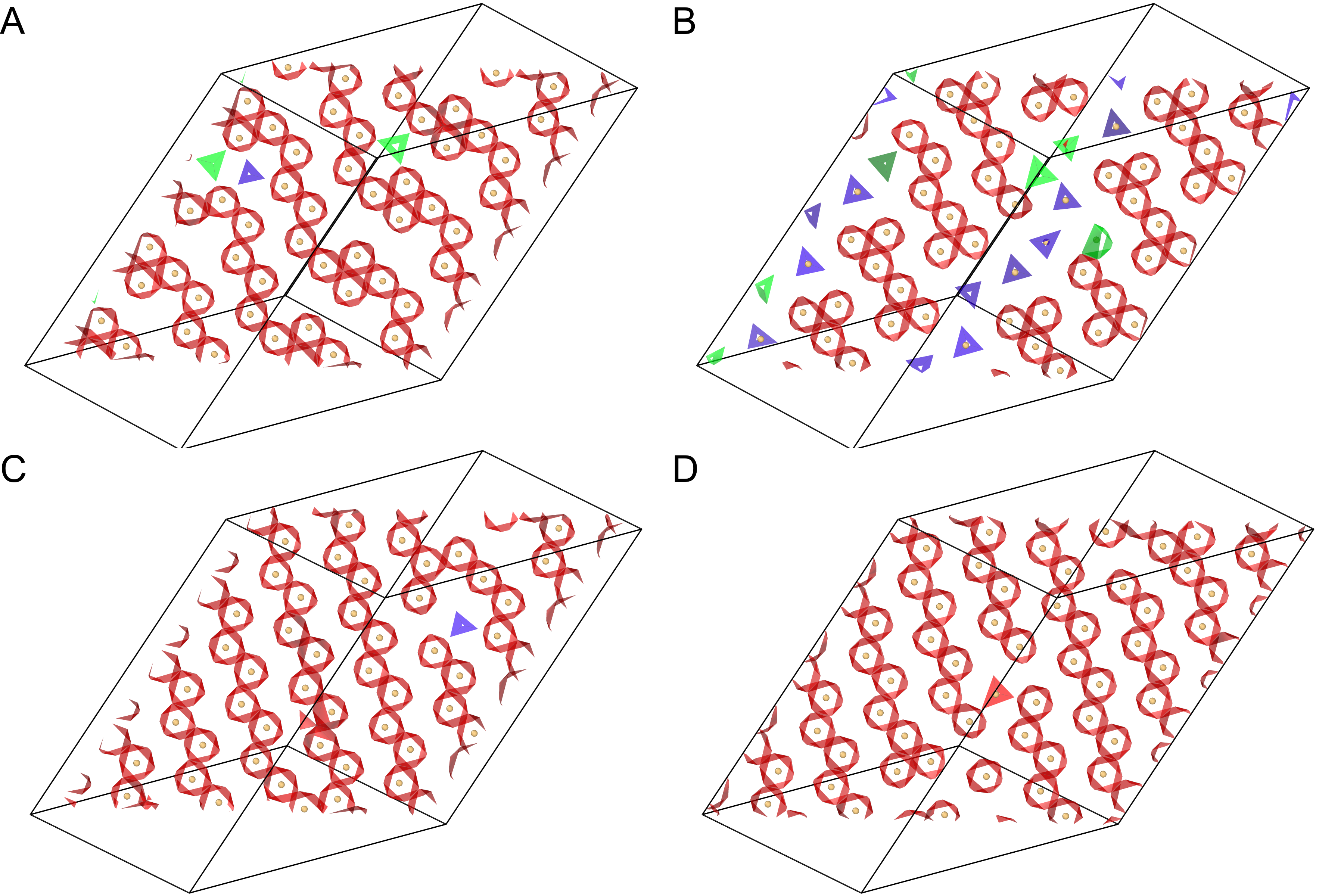}
\caption{\label{fig:HPO_chains} (\textbf{A}, \textbf{B}) Configuration of octahedral chains in two selected planes in the initial HPO phase after formation of the hcp sublattice of O atoms (other planes show similar patterns); (\textbf{C}, \textbf{D}) the same planes after $15.3$ ns of the additional metadynamics run. Note the substantial disorder in the initial chains (\textbf{A}, \textbf{B}) which contain $\alpha$-PbO$_2$-like fragments but do not follow a global periodic pattern. The chains (\textbf{C}, \textbf{D}) clearly show an evolution towards the periodic $2\times2$ pattern.}
\end{figure}

\begin{figure}
\includegraphics[width=8cm]{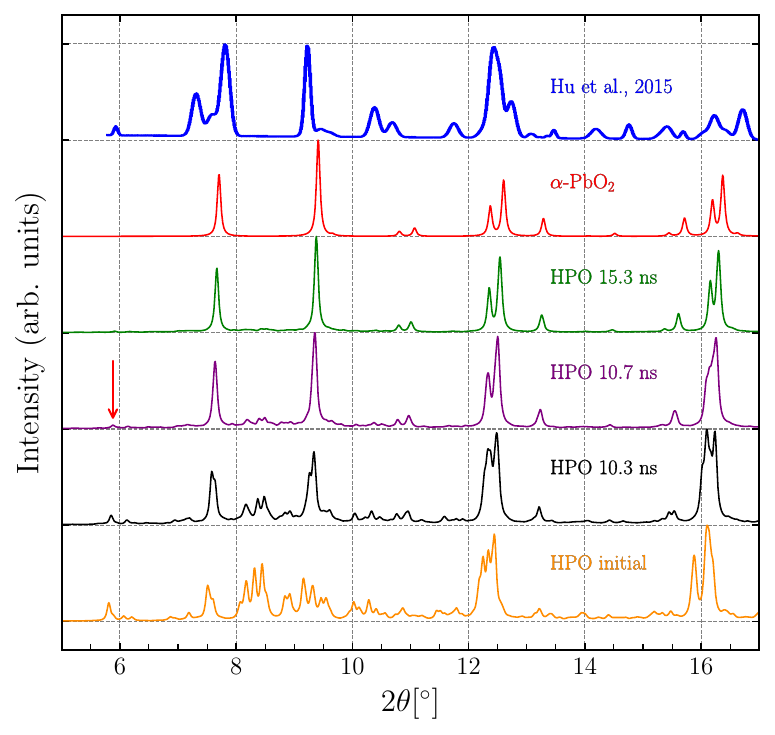}
\caption{\label{fig:HPO_XRD} 
Comparison of the XRD pattern of the initial HPO phase right after the formation of the hcp sublattice of O atoms (Fig.\ref{fig:HPO_chains} A, B) (orange), during the process of ordering of the chains in the additional metadynamics at $p=53$ GPa and 600 K (black, \textcolor{blue}{violet} and green), $\alpha$-PbO$_2$-like phase (red), and the structure found experimentally in Ref.\cite{Hu2015} (blue). The pattern was calculated for the same wavelength $\lambda= 0.4066$ {\AA} as used in Ref.\cite{Hu2015}.
}
\end{figure}

In Fig.\ref{fig:hpo-cv} we show the time evolution of mean Si-O CN, volume, enthalpy of the system and number of 4-, 5- and 6-coordinated Si atoms in the metadynamics run which resulted in the HPO structure. At 300 K the enthalpy can be considered as a very good approximation to the Gibbs free energy of the system. 
The transformation starts at time $3.2$ ns where some Si atoms increase coordination from 4 to 5 (see Fig.\ref{fig:HPO_transition} a). At a later stage (4.3 ns), many Si atoms become 5-coordinated and some start forming octahedra with a coordination of 6 (see Fig.\ref{fig:HPO_transition} b). 
These transformation steps are also well visible in Fig.\ref{fig:hpo-cv} as distinct drops of enthalpy. Finally, around 4.9 ns, the enthalpy drops by nearly 0.2 eV/atom, indicating a dramatic change in the system. 
At this point, the number of 6-coordinated Si atoms increases to over 80 \% (Fig.\ref{fig:hpo-cv}). At the same time, the number of O atoms forming a hcp lattice dramatically jumps from about 10\% to 97 \% (Fig. \ref{fig:hcp-count}), indicating a high degree of ordering of the O sublattice. According to the Ref.\cite{Teter1998}, low enthalpy octahedral structures of SiO$_2$ at high pressures can be represented as hcp sublattice of O atoms where Si atoms fill half of octahedral positions, and octahedra form generic zigzag chains with a $m \times n$ periodicity. Here, we observe a formation of a highly ordered hcp sublattice of O atoms as well as clearly visible chains of Si atoms (Fig.\ref{fig:HPO_chains} a,b). While the chains contain short $\alpha$-PbO$_2$-like fragments, globally they do not follow a regular periodic pattern (Fig.\ref{fig:HPO_chains} A, B). Moreover, we also noted a presence of face-sharing octahedra which represents a violation of the 3rd Pauling rule (similarly to coesite-\RomanNumeralCaps{4}). Therefore, the enthalpy of this structure remains relatively high, about than 0.2 eV/atom above that of the metastable $\alpha$-PbO$_2$-like phase. We thus identified the first step towards creation of an HPO structure, which includes formation of an ordered hcp sublattice of O atoms and chains of octahedra lacking periodicity. For comparison with the XRD pattern found for the HPO structure in Ref.\cite{Hu2015}, we relaxed the structure with the lowest enthalpy (at time 5 ns) to $T=0$ and further compressed it to 53 GPa. 
In Fig.~\ref{fig:HPO_XRD} we compare the XRD pattern of our HPO structure (orange curve) with the one found experimentally\cite{Hu2015} (blue curve) as well as that of the $\alpha$-PbO$_2$-like phase (red curve). Overall, there is a similarity of our pattern with the experimental one. In particular, we note the presence of the peak around 6 deg both in the experimental and our XRD pattern \textcolor{blue}{(see the red arrow in Fig.~\ref{fig:HPO_XRD})}. This peak was also present in the pattern of the HPO structure found in Ref.\cite{Liu2017} while it is not visible in the $\alpha$-PbO$_2$-like phase (see the discussion in Ref.\cite{Liu2017}). In the latter phase, the (100) peak is present at 6 deg, but its intensity is zero due to a vanishing structure factor related to the symmetry of the basis in the ideal structure. Instead, in a defective structure where the chain periodicity is violated, the structure factor becomes non-zero, and the peak becomes visible. Its intensity thus can be regarded as a measure of disorder present in the chains. 

We mention that in one of our metadynamics simulations at 35 GPa and 600 K, a nearly perfect $\alpha$-PbO$_2$-like phase was formed. Here, the formation of hcp sublattice of O atoms was a sudden process, too, while the octahedra chains were created more ordered compared to the lower temperature of 300 K. It can therefore be expected that 
the enthalpy of the structure created at 300 K would further decrease if the chains evolve towards a periodic zig-zag pattern\cite{Teter1998}. This ordering is related to diffusion of Si atoms within the O sublattice. In order to study the structural evolution during this process we performed additional metadynamics simulation with different CVs, targeting the Si-Si coordination, running at the experimental pressure of 53 GPa at which the XRD pattern was measured in Ref.\cite{Hu2015}. \textcolor{blue}{This additional simulation was performed at 300 K, and also at 600 K, in order to accelerate 
the slow diffusion process. It turned out that the process of ordering takes place at 300 K and 600 K in a very similar manner, being significantly faster at 600 K and therefore we further discuss this simulation.} The final HPO structure found in metadynamics at 35 GPa and 300 K was further compressed to 53 GPa and used as starting structure for the new metadynamics using two different Si-Si coordination-related CVs  (the respective switching functions are described in Supp. Mat.). The enthalpy evolution in this simulation is shown in Supp.~Mat. (Fig. 3). After $10.2$ ns, one can see a notable drop of enthalpy of about $0.08$ eV/atom, indicating distinct ordering in the system. Afterwards the enthalpy further continues to decrease, and the lowest value is higher by only $0.12$ eV/atom with respect to the $\alpha$-PbO$_2$-like phase, compared to the initial difference of $0.24$ eV/atom. In Fig.\ref{fig:HPO_chains} C, D one can see the same planes as in the initial state (shown in A, B), after $15.3$ ns. It can be clearly seen that the chains gradually evolve towards the periodic $2\times2$ pattern characteristic for the $\alpha$-PbO$_2$-like phase. 

\textcolor{blue}{This structural evolution is reflected also in the evolution of the respective XRD patterns shown for several times in Fig.~\ref{fig:HPO_XRD}. 
While the structures at 10.3 ns (black curve) and 10.7 ns (violet curve) still display the peak at 6 deg, in the final (lowest enthalpy) one at 15.3 ns (green curve) it is barely visible, since a nearly perfect order of chains was achieved. Overall, the highest similarity to the experimental XRD\cite{Hu2015} is found for the intermediate structure at 10.7 ns where also a minor peak at 7.2 deg, present in the experimental XRD, is still visible. Our results thus suggest that along the pathway towards the $\alpha$-PbO$_2$-like phase, once the hcp sublattice of O atoms is formed, the system evolves via a megabasin of possible structures, differing in the amount of ordering of the Si atoms. Since each of these structures has a different XRD pattern, it would be unrealistic to expect here a perfect quantitative agreement with the experimental data.} Instead, from the above considerations it follows that the experimental structure might represent one of a large variety of intermediate structures along the pathway towards the perfect $\alpha$-PbO$_2$-like phase. We also note that the sublattice of O atoms in the final configuration forms a perfect hcp lattice.

To summarize this subsection, the scenario that emerges from our simulations involves a two-step mechanism. The initial and more easy step is a formation of the hcp sublattice of O atoms. The next, more difficult and slower step is the ordering of Si atoms on the octahedral sites towards a periodic zig-zag pattern, evolving towards the $\alpha$-PbO$_2$-like silica.

\subsection{Pathway to coesite-\RomanNumeralCaps{4} and coesite-\RomanNumeralCaps{5}}

\begin{figure}
\includegraphics[width=8cm]{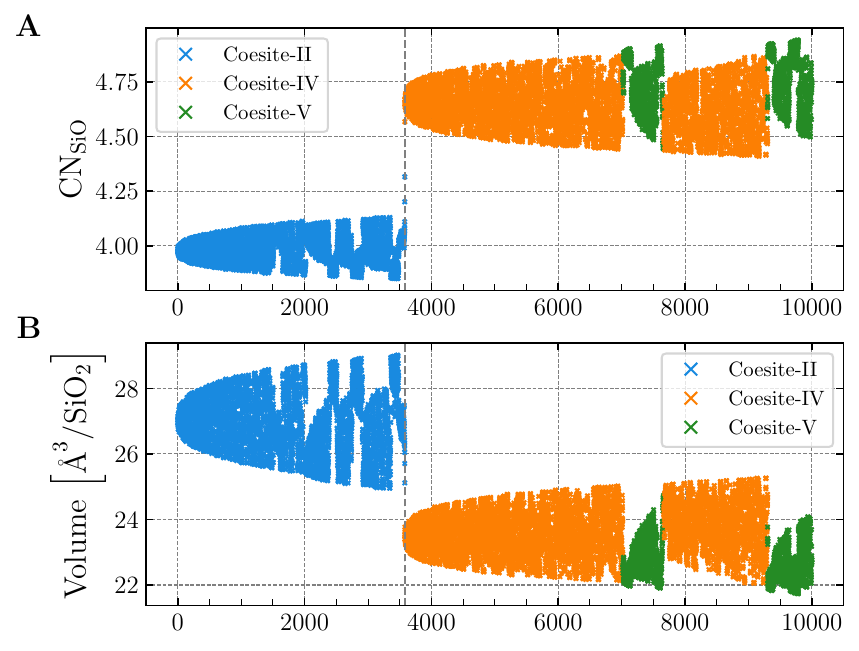}
\includegraphics[width=8cm]{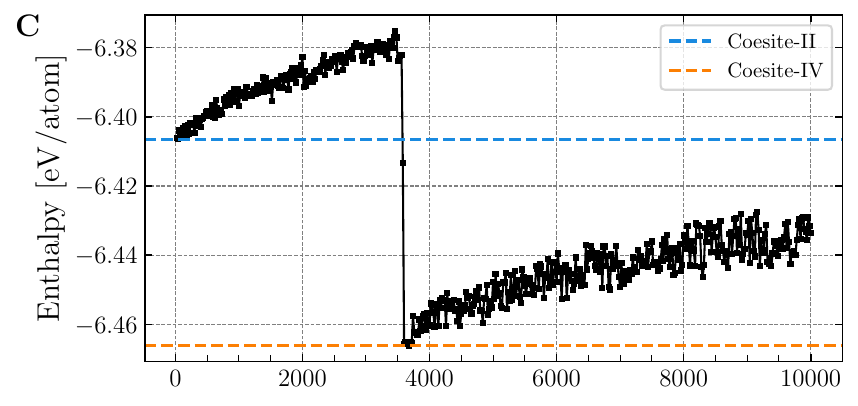}
\includegraphics[width=8cm]{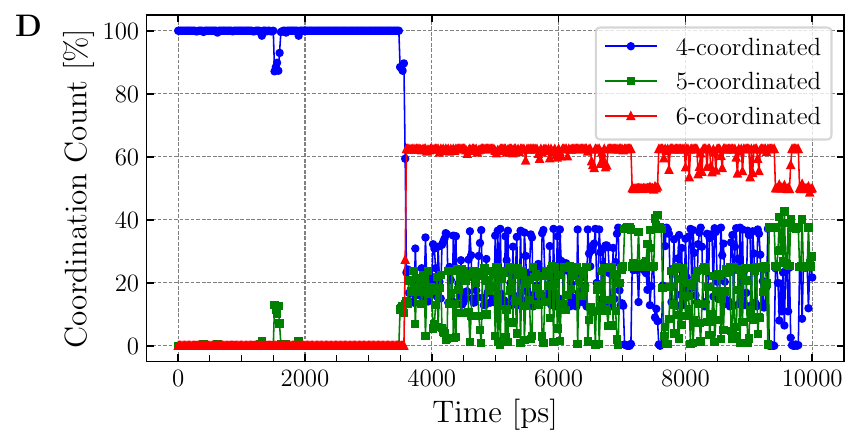}
\caption{\label{fig:coesite4} Evolution of the mean coordination number (\textbf{A}), volume (\textbf{B}), enthalpy (\textbf{C}) and number of 4-, 5- and 6-coordinated Si atoms (\textbf{D}) during the metadynamics simulation at $p=35$ GPa and $T=300$ K which resulted in formation of the coesite-\RomanNumeralCaps{4} and coesite-\RomanNumeralCaps{5}. In the evolution of CVs the initial coesite-\RomanNumeralCaps{2} phase is shown in blue, coesite-\RomanNumeralCaps{4} is shown in orange and the coesite-\RomanNumeralCaps{5} is shown in green.}
\end{figure}

\begin{figure*}
\includegraphics[width=\textwidth]{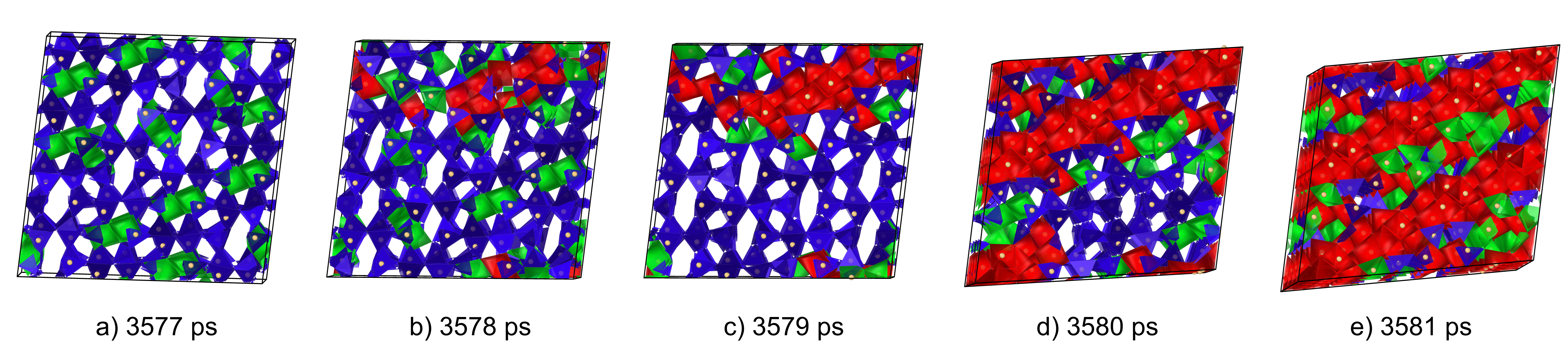}
\caption{\label{fig:coesite4_transition} Evolution of the structure of the system during formation of the coesite-\RomanNumeralCaps{4} at $p=35$ GPa and $T=300$ K. Si atoms with coordination of 4,5 and 6 are shown as blue, green and red polyhedra, respectively.}
\end{figure*}

In Fig.\ref{fig:coesite4} we show the time evolution of mean CN, volume,
enthalpy of the system and number of 4-, 5- and 6-coordinated
Si atoms in the metadynamics run which resulted in the coesite-\RomanNumeralCaps{4}
structure. In this case, the structural evolution appears more straightforward compared to the previous one, since no intermediate structures appear. At $t=3.577$ ns the coordination of Si atoms jumps upwards, accompanied by densification of the system by $19.5$ \% and an enthalpy drop of about 0.06 eV/atom with respect to that of coesite-\RomanNumeralCaps{2}. About 62 \% of Si atoms develop sixfold coordination, while about 25 \% develop a five-fold and 13 \% remains in the original four-fold coordination. The structural evolution is shown in Fig.\ref{fig:coesite4_transition}. The transformation is very fast and occurs in mere 4 ps. 
The structure was relaxed with \textit{ab initio} simulation \cite{Kresse1993, Kresse1996a, Kresse1996b, Kresse1999} and turned out to be identical with coesite-\RomanNumeralCaps{4} found experimentally in Ref.\cite{Bykova2018}. The structure of coesite-\RomanNumeralCaps{4} is unusually complex and is described in detail in Ref.\cite{Bykova2018}. We further simulated a smaller supercell of coesite-\RomanNumeralCaps{4} ($2\times2\times2$ with 384 atoms) by 
plain NPT MD at $T=300$ K and a pressure of 50 GPa where coesite-\RomanNumeralCaps{4} was experimentally found to transform to coesite-\RomanNumeralCaps{5}\cite{Bykova2018}. Within 40 ps of simulation time, we observed a transformation to coesite-\RomanNumeralCaps{5} where two Si atoms increased coordination from 4 to 5. This confirms the very low barrier between the two phases, as noted already in Ref.\cite{Bykova2018}. 

\subsection{Pathway to mixed and disordered structures}

\begin{figure}
\includegraphics[width=8cm]{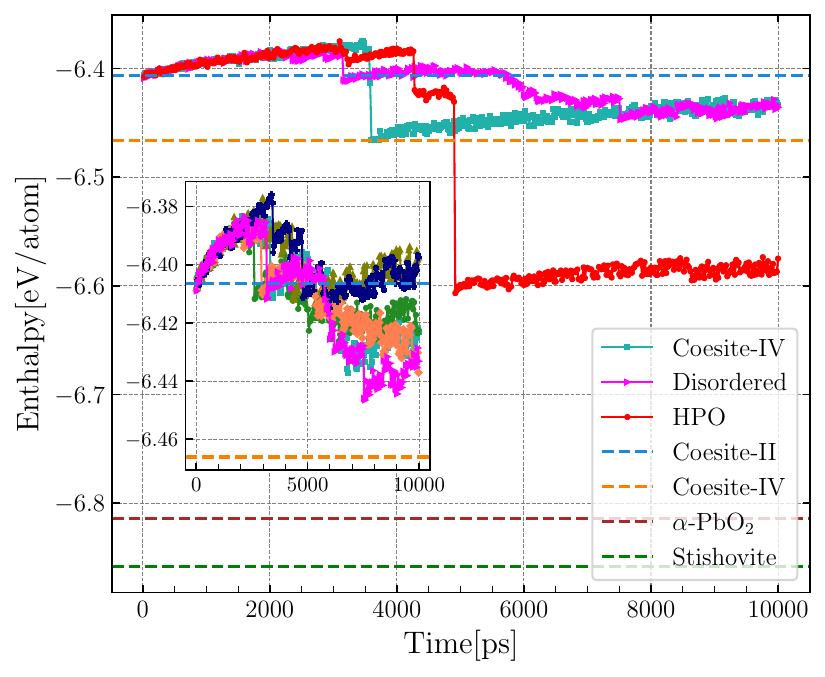}
\caption{\label{fig:enth-comb} Comparison of evolution of the enthalpy during metadynamics simulation at $p=35$ GPa and $T=300$ K, which resulted in HPO, coesite-\RomanNumeralCaps{4} and disordered phases. The inset shows evolution of the enthalpy for other simulations which resulted in mixed or disordered phase.}
\end{figure}

In the inset of Fig.\ref{fig:enth-comb} we show the time evolution of the enthalpy in the simulations  which resulted in the formation of mixed or disordered structures.
These structures are not fully crystalline, and exhibit a coexistence of regions with notably different degree of crystallinity. 
The structures include 4-, 5-, and 6- coordinated Si atoms, with partial crystalline order disturbed by defects. More information about these structures as well as their figures can be found in the Supp. Mat..

\subsection{Discussion}

The results convincingly show the existence of multiple pathways from strongly overpressurized coesite in the region of pressures where its enthalpy is much higher than that of octahedral phases (about 0.45 eV/atom above stishovite or 0.41 eV/atom above the $\alpha$-PbO$_2$-like phase). The pathway towards HPO represents essentially a frustrated attempt to create the $\alpha$-PbO$_2$-like phase, with a non-unique outcome. In this process, the first step, formation of the ordered hcp sublattice of O atoms is the easier one than the second one, ordering of Si atoms, which is difficult due to the complex initial structure of coesite-\RomanNumeralCaps{2}. On the other hand, the other pathway to coesite-\RomanNumeralCaps{4} appears simpler, however, the system remains very high in enthalpy, as discussed already in Ref.\cite{Bykova2018}, \textcolor{blue}{and the volume also remains higher (compare Fig.\ref{fig:hpo-cv} B and Fig.\ref{fig:coesite4} B)}. The Fig.\ref{fig:enth-comb} summarizes the enthalpy evolution of the most important runs at $T=300$ K. As could be expected, the non-crystalline (mixed or disordered) phases are found to have the highest enthalpies, about $0.41$ eV/atom higher than stishovite. Surprisingly, crystalline coesite-\RomanNumeralCaps{4} has only slightly lower enthalpy (higher by 0.39 eV/atom than stishovite), underlining the unusual character of this phase which still bears some similarity to the parent coesite-\RomanNumeralCaps{2} phase. The HPO phase, instead, reaches already in the first step a considerably lower enthalpy ($0.25$ eV/atom than stishovite at 35 GPa) due to more pronounced reconstruction of the parent crystal structure. Its enthalpy continues to further decrease upon the evolution of octahedra chains towards a periodic pattern. Our simulations also point to a clear temperature dependence of the respective pathways. While at 300 K the pathway to coesite-\RomanNumeralCaps{4} and 5 appears preferred, at higher temperature of 600 K the one to HPO is more likely. 

\section{Conclusions}
\label{sec:conc}

We studied the intricate compression mechanism of coesite and found both complex crystalline phases previously observed experimentally, the HPO phase, as well as coesite-\RomanNumeralCaps{4}. We note that we work at pressures close to the experimental pressure where the relevant transitions were observed. This demonstrates that the use of metadynamics based on generic structural variables in combination with accurate ML potential is able to considerably improve the agreement between theory and experiment. The present case represents an unusual one where the structural changes take place under strongly off-equilibrium conditions and the resulting metastable phases may have high enthalpies compared to the equilibrium structure. As noted in Ref.\cite{Bykova2018}, complex metastable phases with high enthalpy are difficult to predict by approaches such as crystal structure prediction, based on the search of all possible metastable structures. We showed here that they  instead can be found by the complementary approach, such as metadynamics, based on simulation of low-energy kinetic pathways accessible from the parent structure at given conditions. We also predict a pronounced temperature dependence of the probability of the realization of the pathways to HPO and to coesite-\RomanNumeralCaps{4} which can be verified experimentally. Last, but not least, our results confirm the high quality of the ACE potential\cite{Erhard2024}.

\section*{Supplementary Material}
The supplementary material provides additional details on the simulations, discussion as well as additional figures.

\begin{acknowledgments}
We emphasize that this work would not have been possible without the landmark achievements of 
Michele Parrinello. All three crucial ingredients, on which this work is based, directly relate to his name - variable-cell MD, metadynamics, and ML potentials. Long ago, the Parrinello-Rahman method pioneered the possibility of simulations of structural phase transitions. Later on, metadynamics and machine learning potentials further greatly enhanced the capabilities of such simulations and brought them much closer to realistic study of kinetic transformation pathways in crystals. R.M. would like to express his deep gratitude to 
Michele Parrinello. Happy birthday, Michele!\\

This work was supported by the Slovak Research and Development Agency under Contract no.~APVV-23-0515 and Grant of Comenius University no. UK/1050/2025. Part of the results was obtained using the computational resources procured in the national project National competence center for high performance computing (project code: 311070AKF2) funded by European Regional Development Fund, EU Structural Funds Informatization of society, Operational Program Integrated Infrastructure. Part of the calculations were performed on the GPU TITAN V provided by the NVIDIA grant.
\end{acknowledgments}

\nocite{Erhard2022}


\begin{thebibliography}{36}%
\makeatletter
\providecommand \@ifxundefined [1]{%
 \@ifx{#1\undefined}
}%
\providecommand \@ifnum [1]{%
 \ifnum #1\expandafter \@firstoftwo
 \else \expandafter \@secondoftwo
 \fi
}%
\providecommand \@ifx [1]{%
 \ifx #1\expandafter \@firstoftwo
 \else \expandafter \@secondoftwo
 \fi
}%
\providecommand \natexlab [1]{#1}%
\providecommand \enquote  [1]{``#1''}%
\providecommand \bibnamefont  [1]{#1}%
\providecommand \bibfnamefont [1]{#1}%
\providecommand \citenamefont [1]{#1}%
\providecommand \href@noop [0]{\@secondoftwo}%
\providecommand \href [0]{\begingroup \@sanitize@url \@href}%
\providecommand \@href[1]{\@@startlink{#1}\@@href}%
\providecommand \@@href[1]{\endgroup#1\@@endlink}%
\providecommand \@sanitize@url [0]{\catcode `\\12\catcode `\$12\catcode
  `\&12\catcode `\#12\catcode `\^12\catcode `\_12\catcode `\%12\relax}%
\providecommand \@@startlink[1]{}%
\providecommand \@@endlink[0]{}%
\providecommand \url  [0]{\begingroup\@sanitize@url \@url }%
\providecommand \@url [1]{\endgroup\@href {#1}{\urlprefix }}%
\providecommand \urlprefix  [0]{URL }%
\providecommand \Eprint [0]{\href }%
\providecommand \doibase [0]{http://dx.doi.org/}%
\providecommand \selectlanguage [0]{\@gobble}%
\providecommand \bibinfo  [0]{\@secondoftwo}%
\providecommand \bibfield  [0]{\@secondoftwo}%
\providecommand \translation [1]{[#1]}%
\providecommand \BibitemOpen [0]{}%
\providecommand \bibitemStop [0]{}%
\providecommand \bibitemNoStop [0]{.\EOS\space}%
\providecommand \EOS [0]{\spacefactor3000\relax}%
\providecommand \BibitemShut  [1]{\csname bibitem#1\endcsname}%
\let\auto@bib@innerbib\@empty
\bibitem [{\citenamefont {Dubrovinsky}\ \emph {et~al.}(2004)\citenamefont
  {Dubrovinsky}, \citenamefont {Dubrovinskaia}, \citenamefont {Prakapenka},
  \citenamefont {Seifert}, \citenamefont {Langenhorst}, \citenamefont
  {Dmitriev}, \citenamefont {Weber},\ and\ \citenamefont {{Le
  Bihan}}}]{dubrovinsky2004}%
  \BibitemOpen
  \bibfield  {author} {\bibinfo {author} {\bibfnamefont {L.}~\bibnamefont
  {Dubrovinsky}}, \bibinfo {author} {\bibfnamefont {N.}~\bibnamefont
  {Dubrovinskaia}}, \bibinfo {author} {\bibfnamefont {V.}~\bibnamefont
  {Prakapenka}}, \bibinfo {author} {\bibfnamefont {F.}~\bibnamefont {Seifert}},
  \bibinfo {author} {\bibfnamefont {F.}~\bibnamefont {Langenhorst}}, \bibinfo
  {author} {\bibfnamefont {V.}~\bibnamefont {Dmitriev}}, \bibinfo {author}
  {\bibfnamefont {H.-P.}\ \bibnamefont {Weber}}, \ and\ \bibinfo {author}
  {\bibfnamefont {T.}~\bibnamefont {{Le Bihan}}},\ }\bibfield  {title}
  {\enquote {\bibinfo {title} {A class of new high-pressure silica
  polymorphs},}\ }\href@noop {} {\bibfield  {journal} {\bibinfo  {journal}
  {Physics of the Earth and Planetary Interiors}\ }\textbf {\bibinfo {volume}
  {143-144}},\ \bibinfo {pages} {231--240} (\bibinfo {year} {2004})},\ \bibinfo
  {note} {new Developments in High-Pressure Mineral Physics and Applications to
  the Earth's Interior}\BibitemShut {NoStop}%
\bibitem [{\citenamefont {Bohlen}\ and\ \citenamefont
  {Boettcher}(1982)}]{Bohlen1982}%
  \BibitemOpen
  \bibfield  {author} {\bibinfo {author} {\bibfnamefont {S.~R.}\ \bibnamefont
  {Bohlen}}\ and\ \bibinfo {author} {\bibfnamefont {A.~L.}\ \bibnamefont
  {Boettcher}},\ }\bibfield  {title} {\enquote {\bibinfo {title} {The quartz
  coesite transformation: A precise determination and the effects of other
  components},}\ }\href {\doibase https://doi.org/10.1029/JB087iB08p07073}
  {\bibfield  {journal} {\bibinfo  {journal} {Journal of Geophysical Research:
  Solid Earth}\ }\textbf {\bibinfo {volume} {87}},\ \bibinfo {pages}
  {7073--7078} (\bibinfo {year} {1982})}\BibitemShut {NoStop}%
\bibitem [{\citenamefont {Kingma}\ \emph
  {et~al.}(1993{\natexlab{a}})\citenamefont {Kingma}, \citenamefont {Hemley},
  \citenamefont {Mao},\ and\ \citenamefont {Veblen}}]{Kingma1993b}%
  \BibitemOpen
  \bibfield  {author} {\bibinfo {author} {\bibfnamefont {K.~J.}\ \bibnamefont
  {Kingma}}, \bibinfo {author} {\bibfnamefont {R.~J.}\ \bibnamefont {Hemley}},
  \bibinfo {author} {\bibfnamefont {H.-k.}\ \bibnamefont {Mao}}, \ and\
  \bibinfo {author} {\bibfnamefont {D.~R.}\ \bibnamefont {Veblen}},\ }\bibfield
   {title} {\enquote {\bibinfo {title} {New high-pressure transformation in
  \ensuremath{\alpha}-quartz},}\ }\href {\doibase 10.1103/PhysRevLett.70.3927}
  {\bibfield  {journal} {\bibinfo  {journal} {Phys. Rev. Lett.}\ }\textbf
  {\bibinfo {volume} {70}},\ \bibinfo {pages} {3927--3930} (\bibinfo {year}
  {1993}{\natexlab{a}})}\BibitemShut {NoStop}%
\bibitem [{\citenamefont {Choudhury}\ and\ \citenamefont
  {Chaplot}(2006)}]{Choudhury2006}%
  \BibitemOpen
  \bibfield  {author} {\bibinfo {author} {\bibfnamefont {N.}~\bibnamefont
  {Choudhury}}\ and\ \bibinfo {author} {\bibfnamefont {S.~L.}\ \bibnamefont
  {Chaplot}},\ }\bibfield  {title} {\enquote {\bibinfo {title} {Ab initio
  studies of phonon softening and high-pressure phase transitions of
  $\ensuremath{\alpha}$-quartz $\mathrm{Si}{\mathrm{o}}_{2}$},}\ }\href
  {\doibase 10.1103/PhysRevB.73.094304} {\bibfield  {journal} {\bibinfo
  {journal} {Phys. Rev. B}\ }\textbf {\bibinfo {volume} {73}},\ \bibinfo
  {pages} {094304} (\bibinfo {year} {2006})}\BibitemShut {NoStop}%
\bibitem [{\citenamefont {Teter}\ \emph {et~al.}(1998)\citenamefont {Teter},
  \citenamefont {Hemley}, \citenamefont {Kresse},\ and\ \citenamefont
  {Hafner}}]{Teter1998}%
  \BibitemOpen
  \bibfield  {author} {\bibinfo {author} {\bibfnamefont {D.~M.}\ \bibnamefont
  {Teter}}, \bibinfo {author} {\bibfnamefont {R.~J.}\ \bibnamefont {Hemley}},
  \bibinfo {author} {\bibfnamefont {G.}~\bibnamefont {Kresse}}, \ and\ \bibinfo
  {author} {\bibfnamefont {J.}~\bibnamefont {Hafner}},\ }\bibfield  {title}
  {\enquote {\bibinfo {title} {High pressure polymorphism in silica},}\ }\href
  {\doibase 10.1103/PhysRevLett.80.2145} {\bibfield  {journal} {\bibinfo
  {journal} {Phys. Rev. Lett.}\ }\textbf {\bibinfo {volume} {80}},\ \bibinfo
  {pages} {2145--2148} (\bibinfo {year} {1998})}\BibitemShut {NoStop}%
\bibitem [{\citenamefont {Hemley}\ \emph {et~al.}(1988)\citenamefont {Hemley},
  \citenamefont {Jephcoat}, \citenamefont {Mao}, \citenamefont {Ming},\ and\
  \citenamefont {Manghnani}}]{Hemley1988}%
  \BibitemOpen
  \bibfield  {author} {\bibinfo {author} {\bibfnamefont {R.~J.}\ \bibnamefont
  {Hemley}}, \bibinfo {author} {\bibfnamefont {A.~P.}\ \bibnamefont
  {Jephcoat}}, \bibinfo {author} {\bibfnamefont {H.~K.}\ \bibnamefont {Mao}},
  \bibinfo {author} {\bibfnamefont {L.~C.}\ \bibnamefont {Ming}}, \ and\
  \bibinfo {author} {\bibfnamefont {M.~H.}\ \bibnamefont {Manghnani}},\
  }\bibfield  {title} {\enquote {\bibinfo {title} {Pressure-induced
  amorphization of crystalline silica},}\ }\href {\doibase 10.1038/334052a0}
  {\bibfield  {journal} {\bibinfo  {journal} {Nature}\ }\textbf {\bibinfo
  {volume} {334}},\ \bibinfo {pages} {52--54} (\bibinfo {year}
  {1988})}\BibitemShut {NoStop}%
\bibitem [{\citenamefont {Kingma}\ \emph
  {et~al.}(1993{\natexlab{b}})\citenamefont {Kingma}, \citenamefont {Hemley},
  \citenamefont {Mao},\ and\ \citenamefont {Veblen}}]{Kingma1993a}%
  \BibitemOpen
  \bibfield  {author} {\bibinfo {author} {\bibfnamefont {K.~J.}\ \bibnamefont
  {Kingma}}, \bibinfo {author} {\bibfnamefont {R.~J.}\ \bibnamefont {Hemley}},
  \bibinfo {author} {\bibfnamefont {H.-k.}\ \bibnamefont {Mao}}, \ and\
  \bibinfo {author} {\bibfnamefont {D.~R.}\ \bibnamefont {Veblen}},\ }\bibfield
   {title} {\enquote {\bibinfo {title} {New high-pressure transformation in
  \ensuremath{\alpha}-quartz},}\ }\href {\doibase 10.1103/PhysRevLett.70.3927}
  {\bibfield  {journal} {\bibinfo  {journal} {Phys. Rev. Lett.}\ }\textbf
  {\bibinfo {volume} {70}},\ \bibinfo {pages} {3927--3930} (\bibinfo {year}
  {1993}{\natexlab{b}})}\BibitemShut {NoStop}%
\bibitem [{\citenamefont {Laio}\ and\ \citenamefont
  {Parrinello}(2002)}]{Laio2002}%
  \BibitemOpen
  \bibfield  {author} {\bibinfo {author} {\bibfnamefont {A.}~\bibnamefont
  {Laio}}\ and\ \bibinfo {author} {\bibfnamefont {M.}~\bibnamefont
  {Parrinello}},\ }\bibfield  {title} {\enquote {\bibinfo {title} {Escaping
  free-energy minima},}\ }\href {\doibase 10.1073/pnas.202427399} {\bibfield
  {journal} {\bibinfo  {journal} {Proceedings of the National Academy of
  Sciences}\ }\textbf {\bibinfo {volume} {99}},\ \bibinfo {pages}
  {12562--12566} (\bibinfo {year} {2002})}\BibitemShut {NoStop}%
\bibitem [{\citenamefont {Martoňák}\ \emph {et~al.}(2006)\citenamefont
  {Martoňák}, \citenamefont {Donadio}, \citenamefont {Oganov},\ and\
  \citenamefont {Parrinello}}]{Martonak2006}%
  \BibitemOpen
  \bibfield  {author} {\bibinfo {author} {\bibfnamefont {R.}~\bibnamefont
  {Martoňák}}, \bibinfo {author} {\bibfnamefont {D.}~\bibnamefont {Donadio}},
  \bibinfo {author} {\bibfnamefont {A.~R.}\ \bibnamefont {Oganov}}, \ and\
  \bibinfo {author} {\bibfnamefont {M.}~\bibnamefont {Parrinello}},\ }\bibfield
   {title} {\enquote {\bibinfo {title} {Crystal structure transformations in
  sio2 from classical and ab initio metadynamics},}\ }\href {\doibase
  10.1038/nmat1696} {\bibfield  {journal} {\bibinfo  {journal} {Nature
  Materials}\ }\textbf {\bibinfo {volume} {5}},\ \bibinfo {pages} {623--626}
  (\bibinfo {year} {2006})}\BibitemShut {NoStop}%
\bibitem [{\citenamefont {Marto\ifmmode~\check{n}\else \v{n}\fi{}\'ak}\ \emph
  {et~al.}(2007)\citenamefont {Marto\ifmmode~\check{n}\else \v{n}\fi{}\'ak},
  \citenamefont {Donadio}, \citenamefont {Oganov},\ and\ \citenamefont
  {Parrinello}}]{Martonak2007}%
  \BibitemOpen
  \bibfield  {author} {\bibinfo {author} {\bibfnamefont {R.}~\bibnamefont
  {Marto\ifmmode~\check{n}\else \v{n}\fi{}\'ak}}, \bibinfo {author}
  {\bibfnamefont {D.}~\bibnamefont {Donadio}}, \bibinfo {author} {\bibfnamefont
  {A.~R.}\ \bibnamefont {Oganov}}, \ and\ \bibinfo {author} {\bibfnamefont
  {M.}~\bibnamefont {Parrinello}},\ }\bibfield  {title} {\enquote {\bibinfo
  {title} {From four- to six-coordinated silica: Transformation pathways from
  metadynamics},}\ }\href {\doibase 10.1103/PhysRevB.76.014120} {\bibfield
  {journal} {\bibinfo  {journal} {Phys. Rev. B}\ }\textbf {\bibinfo {volume}
  {76}},\ \bibinfo {pages} {014120} (\bibinfo {year} {2007})}\BibitemShut
  {NoStop}%
\bibitem [{\citenamefont {Černok}\ \emph
  {et~al.}(2014{\natexlab{a}})\citenamefont {Černok}, \citenamefont
  {Ballaran}, \citenamefont {Caracas}, \citenamefont {Miyajima}, \citenamefont
  {Bykova}, \citenamefont {Prakapenka}, \citenamefont {Liermann},\ and\
  \citenamefont {Dubrovinsky}}]{Cernok2014a}%
  \BibitemOpen
  \bibfield  {author} {\bibinfo {author} {\bibfnamefont {A.}~\bibnamefont
  {Černok}}, \bibinfo {author} {\bibfnamefont {T.~B.}\ \bibnamefont
  {Ballaran}}, \bibinfo {author} {\bibfnamefont {R.}~\bibnamefont {Caracas}},
  \bibinfo {author} {\bibfnamefont {N.}~\bibnamefont {Miyajima}}, \bibinfo
  {author} {\bibfnamefont {E.}~\bibnamefont {Bykova}}, \bibinfo {author}
  {\bibfnamefont {V.}~\bibnamefont {Prakapenka}}, \bibinfo {author}
  {\bibfnamefont {H.-P.}\ \bibnamefont {Liermann}}, \ and\ \bibinfo {author}
  {\bibfnamefont {L.}~\bibnamefont {Dubrovinsky}},\ }\bibfield  {title}
  {\enquote {\bibinfo {title} {Pressure-induced phase transitions in
  coesite},}\ }\href {\doibase doi:10.2138/am.2014.4585} {\bibfield  {journal}
  {\bibinfo  {journal} {American Mineralogist}\ }\textbf {\bibinfo {volume}
  {99}},\ \bibinfo {pages} {755--763} (\bibinfo {year}
  {2014}{\natexlab{a}})}\BibitemShut {NoStop}%
\bibitem [{\citenamefont {Černok}\ \emph
  {et~al.}(2014{\natexlab{b}})\citenamefont {Černok}, \citenamefont {Bykova},
  \citenamefont {Ballaran}, \citenamefont {Liermann}, \citenamefont
  {Hanfland},\ and\ \citenamefont {Dubrovinsky}}]{Cernok2014b}%
  \BibitemOpen
  \bibfield  {author} {\bibinfo {author} {\bibfnamefont {A.}~\bibnamefont
  {Černok}}, \bibinfo {author} {\bibfnamefont {E.}~\bibnamefont {Bykova}},
  \bibinfo {author} {\bibfnamefont {T.~B.}\ \bibnamefont {Ballaran}}, \bibinfo
  {author} {\bibfnamefont {H.-P.}\ \bibnamefont {Liermann}}, \bibinfo {author}
  {\bibfnamefont {M.}~\bibnamefont {Hanfland}}, \ and\ \bibinfo {author}
  {\bibfnamefont {L.}~\bibnamefont {Dubrovinsky}},\ }\bibfield  {title}
  {\enquote {\bibinfo {title} {High-pressure crystal chemistry of coesite-i and
  its transition to coesite-ii},}\ }\href {\doibase doi:10.1515/zkri-2014-1763}
  {\bibfield  {journal} {\bibinfo  {journal} {Zeitschrift für Kristallographie
  - Crystalline Materials}\ }\textbf {\bibinfo {volume} {229}},\ \bibinfo
  {pages} {761--773} (\bibinfo {year} {2014}{\natexlab{b}})}\BibitemShut
  {NoStop}%
\bibitem [{\citenamefont {Wu}\ \emph {et~al.}(2018)\citenamefont {Wu},
  \citenamefont {Liu}, \citenamefont {Huang}, \citenamefont {Fei},
  \citenamefont {Feng},\ and\ \citenamefont {Redfern}}]{Wu2018}%
  \BibitemOpen
  \bibfield  {author} {\bibinfo {author} {\bibfnamefont {Y.}~\bibnamefont
  {Wu}}, \bibinfo {author} {\bibfnamefont {H.}~\bibnamefont {Liu}}, \bibinfo
  {author} {\bibfnamefont {H.}~\bibnamefont {Huang}}, \bibinfo {author}
  {\bibfnamefont {Y.}~\bibnamefont {Fei}}, \bibinfo {author} {\bibfnamefont
  {X.}~\bibnamefont {Feng}}, \ and\ \bibinfo {author} {\bibfnamefont
  {S.~A.~T.}\ \bibnamefont {Redfern}},\ }\bibfield  {title} {\enquote {\bibinfo
  {title} {Pressure-induced structural modulations in coesite},}\ }\href
  {\doibase 10.1103/PhysRevB.98.104106} {\bibfield  {journal} {\bibinfo
  {journal} {Phys. Rev. B}\ }\textbf {\bibinfo {volume} {98}},\ \bibinfo
  {pages} {104106} (\bibinfo {year} {2018})}\BibitemShut {NoStop}%
\bibitem [{\citenamefont {Hu}\ \emph {et~al.}(2015)\citenamefont {Hu},
  \citenamefont {Shu}, \citenamefont {Cadien}, \citenamefont {Meng},
  \citenamefont {Yang}, \citenamefont {Sheng},\ and\ \citenamefont
  {Mao}}]{Hu2015}%
  \BibitemOpen
  \bibfield  {author} {\bibinfo {author} {\bibfnamefont {Q.~Y.}\ \bibnamefont
  {Hu}}, \bibinfo {author} {\bibfnamefont {J.-F.}\ \bibnamefont {Shu}},
  \bibinfo {author} {\bibfnamefont {A.}~\bibnamefont {Cadien}}, \bibinfo
  {author} {\bibfnamefont {Y.}~\bibnamefont {Meng}}, \bibinfo {author}
  {\bibfnamefont {W.~G.}\ \bibnamefont {Yang}}, \bibinfo {author}
  {\bibfnamefont {H.~W.}\ \bibnamefont {Sheng}}, \ and\ \bibinfo {author}
  {\bibfnamefont {H.-K.}\ \bibnamefont {Mao}},\ }\bibfield  {title} {\enquote
  {\bibinfo {title} {Polymorphic phase transition mechanism of compressed
  coesite},}\ }\href {\doibase 10.1038/ncomms7630} {\bibfield  {journal}
  {\bibinfo  {journal} {Nature Communications}\ }\textbf {\bibinfo {volume}
  {6}},\ \bibinfo {pages} {6630} (\bibinfo {year} {2015})}\BibitemShut
  {NoStop}%
\bibitem [{\citenamefont {Liu}\ \emph {et~al.}(2017)\citenamefont {Liu},
  \citenamefont {Wu}, \citenamefont {Liang}, \citenamefont {Liu}, \citenamefont
  {Miranda},\ and\ \citenamefont {Scandolo}}]{Liu2017}%
  \BibitemOpen
  \bibfield  {author} {\bibinfo {author} {\bibfnamefont {W.}~\bibnamefont
  {Liu}}, \bibinfo {author} {\bibfnamefont {X.}~\bibnamefont {Wu}}, \bibinfo
  {author} {\bibfnamefont {Y.}~\bibnamefont {Liang}}, \bibinfo {author}
  {\bibfnamefont {C.}~\bibnamefont {Liu}}, \bibinfo {author} {\bibfnamefont
  {C.~R.}\ \bibnamefont {Miranda}}, \ and\ \bibinfo {author} {\bibfnamefont
  {S.}~\bibnamefont {Scandolo}},\ }\bibfield  {title} {\enquote {\bibinfo
  {title} {Multiple pathways in pressure-induced phase transition of
  coesite},}\ }\href {\doibase 10.1073/pnas.1710651114} {\bibfield  {journal}
  {\bibinfo  {journal} {Proceedings of the National Academy of Sciences}\
  }\textbf {\bibinfo {volume} {114}},\ \bibinfo {pages} {12894--12899}
  (\bibinfo {year} {2017})}\BibitemShut {NoStop}%
\bibitem [{\citenamefont {Bykova}\ \emph {et~al.}(2018)\citenamefont {Bykova},
  \citenamefont {Bykov}, \citenamefont {{\v{C}}ernok}, \citenamefont {Tidholm},
  \citenamefont {Simak}, \citenamefont {Hellman}, \citenamefont {Belov},
  \citenamefont {Abrikosov}, \citenamefont {Liermann}, \citenamefont
  {Hanfland}, \citenamefont {Prakapenka}, \citenamefont {Prescher},
  \citenamefont {Dubrovinskaia},\ and\ \citenamefont
  {Dubrovinsky}}]{Bykova2018}%
  \BibitemOpen
  \bibfield  {author} {\bibinfo {author} {\bibfnamefont {E.}~\bibnamefont
  {Bykova}}, \bibinfo {author} {\bibfnamefont {M.}~\bibnamefont {Bykov}},
  \bibinfo {author} {\bibfnamefont {A.}~\bibnamefont {{\v{C}}ernok}}, \bibinfo
  {author} {\bibfnamefont {J.}~\bibnamefont {Tidholm}}, \bibinfo {author}
  {\bibfnamefont {S.~I.}\ \bibnamefont {Simak}}, \bibinfo {author}
  {\bibfnamefont {O.}~\bibnamefont {Hellman}}, \bibinfo {author} {\bibfnamefont
  {M.~P.}\ \bibnamefont {Belov}}, \bibinfo {author} {\bibfnamefont {I.~A.}\
  \bibnamefont {Abrikosov}}, \bibinfo {author} {\bibfnamefont {H.-P.}\
  \bibnamefont {Liermann}}, \bibinfo {author} {\bibfnamefont {M.}~\bibnamefont
  {Hanfland}}, \bibinfo {author} {\bibfnamefont {V.~B.}\ \bibnamefont
  {Prakapenka}}, \bibinfo {author} {\bibfnamefont {C.}~\bibnamefont
  {Prescher}}, \bibinfo {author} {\bibfnamefont {N.}~\bibnamefont
  {Dubrovinskaia}}, \ and\ \bibinfo {author} {\bibfnamefont {L.}~\bibnamefont
  {Dubrovinsky}},\ }\bibfield  {title} {\enquote {\bibinfo {title} {Metastable
  silica high pressure polymorphs as structural proxies of deep earth silicate
  melts},}\ }\href {\doibase 10.1038/s41467-018-07265-z} {\bibfield  {journal}
  {\bibinfo  {journal} {Nature Communications}\ }\textbf {\bibinfo {volume}
  {9}},\ \bibinfo {pages} {4789} (\bibinfo {year} {2018})}\BibitemShut
  {NoStop}%
\bibitem [{\citenamefont {Barducci}, \citenamefont {Bonomi},\ and\
  \citenamefont {Parrinello}(2011)}]{Barducci2011}%
  \BibitemOpen
  \bibfield  {author} {\bibinfo {author} {\bibfnamefont {A.}~\bibnamefont
  {Barducci}}, \bibinfo {author} {\bibfnamefont {M.}~\bibnamefont {Bonomi}}, \
  and\ \bibinfo {author} {\bibfnamefont {M.}~\bibnamefont {Parrinello}},\
  }\bibfield  {title} {\enquote {\bibinfo {title} {Metadynamics},}\ }\href
  {\doibase https://doi.org/10.1002/wcms.31} {\bibfield  {journal} {\bibinfo
  {journal} {WIREs Computational Molecular Science}\ }\textbf {\bibinfo
  {volume} {1}},\ \bibinfo {pages} {826--843} (\bibinfo {year}
  {2011})}\BibitemShut {NoStop}%
\bibitem [{\citenamefont {Bussi}\ and\ \citenamefont {Laio}(2020)}]{Bussi2020}%
  \BibitemOpen
  \bibfield  {author} {\bibinfo {author} {\bibfnamefont {G.}~\bibnamefont
  {Bussi}}\ and\ \bibinfo {author} {\bibfnamefont {A.}~\bibnamefont {Laio}},\
  }\bibfield  {title} {\enquote {\bibinfo {title} {Using metadynamics to
  explore complex free-energy landscapes},}\ }\href {\doibase
  10.1038/s42254-020-0153-0} {\bibfield  {journal} {\bibinfo  {journal} {Nature
  Reviews Physics}\ }\textbf {\bibinfo {volume} {2}},\ \bibinfo {pages}
  {200--212} (\bibinfo {year} {2020})}\BibitemShut {NoStop}%
\bibitem [{\citenamefont {Badin}\ and\ \citenamefont
  {Martoňák}(2021)}]{Badin2021}%
  \BibitemOpen
  \bibfield  {author} {\bibinfo {author} {\bibfnamefont {M.}~\bibnamefont
  {Badin}}\ and\ \bibinfo {author} {\bibfnamefont {R.}~\bibnamefont
  {Martoňák}},\ }\bibfield  {title} {\enquote {\bibinfo {title} {Nucleating a
  different coordination in a crystal under pressure: A study of the $b1-b2$
  transition in nacl by metadynamics},}\ }\href {\doibase
  10.1103/PhysRevLett.127.105701} {\bibfield  {journal} {\bibinfo  {journal}
  {Phys. Rev. Lett.}\ }\textbf {\bibinfo {volume} {127}},\ \bibinfo {pages}
  {105701} (\bibinfo {year} {2021})}\BibitemShut {NoStop}%
\bibitem [{\citenamefont {Thompson}\ \emph {et~al.}(2022)\citenamefont
  {Thompson}, \citenamefont {Aktulga}, \citenamefont {Berger}, \citenamefont
  {Bolintineanu}, \citenamefont {Brown}, \citenamefont {Crozier}, \citenamefont
  {in~'t Veld}, \citenamefont {Kohlmeyer}, \citenamefont {Moore}, \citenamefont
  {Nguyen}, \citenamefont {Shan}, \citenamefont {Stevens}, \citenamefont
  {Tranchida}, \citenamefont {Trott},\ and\ \citenamefont {Plimpton}}]{LAMMPS}%
  \BibitemOpen
  \bibfield  {author} {\bibinfo {author} {\bibfnamefont {A.~P.}\ \bibnamefont
  {Thompson}}, \bibinfo {author} {\bibfnamefont {H.~M.}\ \bibnamefont
  {Aktulga}}, \bibinfo {author} {\bibfnamefont {R.}~\bibnamefont {Berger}},
  \bibinfo {author} {\bibfnamefont {D.~S.}\ \bibnamefont {Bolintineanu}},
  \bibinfo {author} {\bibfnamefont {W.~M.}\ \bibnamefont {Brown}}, \bibinfo
  {author} {\bibfnamefont {P.~S.}\ \bibnamefont {Crozier}}, \bibinfo {author}
  {\bibfnamefont {P.~J.}\ \bibnamefont {in~'t Veld}}, \bibinfo {author}
  {\bibfnamefont {A.}~\bibnamefont {Kohlmeyer}}, \bibinfo {author}
  {\bibfnamefont {S.~G.}\ \bibnamefont {Moore}}, \bibinfo {author}
  {\bibfnamefont {T.~D.}\ \bibnamefont {Nguyen}}, \bibinfo {author}
  {\bibfnamefont {R.}~\bibnamefont {Shan}}, \bibinfo {author} {\bibfnamefont
  {M.~J.}\ \bibnamefont {Stevens}}, \bibinfo {author} {\bibfnamefont
  {J.}~\bibnamefont {Tranchida}}, \bibinfo {author} {\bibfnamefont
  {C.}~\bibnamefont {Trott}}, \ and\ \bibinfo {author} {\bibfnamefont {S.~J.}\
  \bibnamefont {Plimpton}},\ }\bibfield  {title} {\enquote {\bibinfo {title}
  {{LAMMPS} - a flexible simulation tool for particle-based materials modeling
  at the atomic, meso, and continuum scales},}\ }\href {\doibase
  10.1016/j.cpc.2021.108171} {\bibfield  {journal} {\bibinfo  {journal} {Comp.
  Phys. Comm.}\ }\textbf {\bibinfo {volume} {271}},\ \bibinfo {pages} {108171}
  (\bibinfo {year} {2022})}\BibitemShut {NoStop}%
\bibitem [{\citenamefont {Tribello}\ \emph {et~al.}(2014)\citenamefont
  {Tribello}, \citenamefont {Bonomi}, \citenamefont {Branduardi}, \citenamefont
  {Camilloni},\ and\ \citenamefont {Bussi}}]{PLUMED}%
  \BibitemOpen
  \bibfield  {author} {\bibinfo {author} {\bibfnamefont {G.~A.}\ \bibnamefont
  {Tribello}}, \bibinfo {author} {\bibfnamefont {M.}~\bibnamefont {Bonomi}},
  \bibinfo {author} {\bibfnamefont {D.}~\bibnamefont {Branduardi}}, \bibinfo
  {author} {\bibfnamefont {C.}~\bibnamefont {Camilloni}}, \ and\ \bibinfo
  {author} {\bibfnamefont {G.}~\bibnamefont {Bussi}},\ }\bibfield  {title}
  {\enquote {\bibinfo {title} {{PLUMED} 2: New feathers for an old bird},}\
  }\href {\doibase 10.1016/j.cpc.2013.09.018} {\bibfield  {journal} {\bibinfo
  {journal} {Computer Physics Communications}\ }\textbf {\bibinfo {volume}
  {185}},\ \bibinfo {pages} {604--613} (\bibinfo {year} {2014})}\BibitemShut
  {NoStop}%
\bibitem [{\citenamefont {Nosé}(1984)}]{Nose1984}%
  \BibitemOpen
  \bibfield  {author} {\bibinfo {author} {\bibfnamefont {S.}~\bibnamefont
  {Nosé}},\ }\bibfield  {title} {\enquote {\bibinfo {title} {A unified
  formulation of the constant temperature molecular dynamics methods},}\ }\href
  {\doibase 10.1063/1.447334} {\bibfield  {journal} {\bibinfo  {journal} {The
  Journal of Chemical Physics}\ }\textbf {\bibinfo {volume} {81}},\ \bibinfo
  {pages} {511--519} (\bibinfo {year} {1984})}\BibitemShut {NoStop}%
\bibitem [{\citenamefont {Hoover}(1985)}]{Hoover1985}%
  \BibitemOpen
  \bibfield  {author} {\bibinfo {author} {\bibfnamefont {W.~G.}\ \bibnamefont
  {Hoover}},\ }\bibfield  {title} {\enquote {\bibinfo {title} {Canonical
  dynamics: Equilibrium phase-space distributions},}\ }\href {\doibase
  10.1103/PhysRevA.31.1695} {\bibfield  {journal} {\bibinfo  {journal} {Phys.
  Rev. A}\ }\textbf {\bibinfo {volume} {31}},\ \bibinfo {pages} {1695--1697}
  (\bibinfo {year} {1985})}\BibitemShut {NoStop}%
\bibitem [{\citenamefont {Parrinello}\ and\ \citenamefont
  {Rahman}(1981)}]{Parrinello1981}%
  \BibitemOpen
  \bibfield  {author} {\bibinfo {author} {\bibfnamefont {M.}~\bibnamefont
  {Parrinello}}\ and\ \bibinfo {author} {\bibfnamefont {A.}~\bibnamefont
  {Rahman}},\ }\bibfield  {title} {\enquote {\bibinfo {title} {Polymorphic
  transitions in single crystals: A new molecular dynamics method},}\ }\href
  {\doibase 10.1063/1.328693} {\bibfield  {journal} {\bibinfo  {journal}
  {Journal of Applied Physics}\ }\textbf {\bibinfo {volume} {52}},\ \bibinfo
  {pages} {7182--7190} (\bibinfo {year} {1981})}\BibitemShut {NoStop}%
\bibitem [{\citenamefont {Erhard}\ \emph {et~al.}(2024)\citenamefont {Erhard},
  \citenamefont {Rohrer}, \citenamefont {Albe},\ and\ \citenamefont
  {Deringer}}]{Erhard2024}%
  \BibitemOpen
  \bibfield  {author} {\bibinfo {author} {\bibfnamefont {L.~C.}\ \bibnamefont
  {Erhard}}, \bibinfo {author} {\bibfnamefont {J.}~\bibnamefont {Rohrer}},
  \bibinfo {author} {\bibfnamefont {K.}~\bibnamefont {Albe}}, \ and\ \bibinfo
  {author} {\bibfnamefont {V.~L.}\ \bibnamefont {Deringer}},\ }\bibfield
  {title} {\enquote {\bibinfo {title} {Modelling atomic and nanoscale structure
  in the silicon--oxygen system through active machine learning},}\ }\href
  {\doibase 10.1038/s41467-024-45840-9} {\bibfield  {journal} {\bibinfo
  {journal} {Nature Communications}\ }\textbf {\bibinfo {volume} {15}},\
  \bibinfo {pages} {1927} (\bibinfo {year} {2024})}\BibitemShut {NoStop}%
\bibitem [{\citenamefont {Drautz}(2019)}]{ACE_Drautz2019}%
  \BibitemOpen
  \bibfield  {author} {\bibinfo {author} {\bibfnamefont {R.}~\bibnamefont
  {Drautz}},\ }\bibfield  {title} {\enquote {\bibinfo {title} {Atomic cluster
  expansion for accurate and transferable interatomic potentials},}\ }\href
  {\doibase 10.1103/PhysRevB.99.014104} {\bibfield  {journal} {\bibinfo
  {journal} {Phys. Rev. B}\ }\textbf {\bibinfo {volume} {99}},\ \bibinfo
  {pages} {014104} (\bibinfo {year} {2019})}\BibitemShut {NoStop}%
\bibitem [{\citenamefont {Kresse}\ and\ \citenamefont
  {Hafner}(1993)}]{Kresse1993}%
  \BibitemOpen
  \bibfield  {author} {\bibinfo {author} {\bibfnamefont {G.}~\bibnamefont
  {Kresse}}\ and\ \bibinfo {author} {\bibfnamefont {J.}~\bibnamefont
  {Hafner}},\ }\bibfield  {title} {\enquote {\bibinfo {title} {Ab initio
  molecular dynamics for liquid metals},}\ }\href {\doibase
  10.1103/PhysRevB.47.558} {\bibfield  {journal} {\bibinfo  {journal} {Phys.
  Rev. B}\ }\textbf {\bibinfo {volume} {47}},\ \bibinfo {pages} {558--561}
  (\bibinfo {year} {1993})}\BibitemShut {NoStop}%
\bibitem [{\citenamefont {Kresse}\ and\ \citenamefont
  {Furthm\"uller}(1996)}]{Kresse1996a}%
  \BibitemOpen
  \bibfield  {author} {\bibinfo {author} {\bibfnamefont {G.}~\bibnamefont
  {Kresse}}\ and\ \bibinfo {author} {\bibfnamefont {J.}~\bibnamefont
  {Furthm\"uller}},\ }\bibfield  {title} {\enquote {\bibinfo {title} {Efficient
  iterative schemes for ab initio total-energy calculations using a plane-wave
  basis set},}\ }\href {\doibase 10.1103/PhysRevB.54.11169} {\bibfield
  {journal} {\bibinfo  {journal} {Phys. Rev. B}\ }\textbf {\bibinfo {volume}
  {54}},\ \bibinfo {pages} {11169--11186} (\bibinfo {year} {1996})}\BibitemShut
  {NoStop}%
\bibitem [{\citenamefont {Kresse}\ and\ \citenamefont
  {Furthmüller}(1996)}]{Kresse1996b}%
  \BibitemOpen
  \bibfield  {author} {\bibinfo {author} {\bibfnamefont {G.}~\bibnamefont
  {Kresse}}\ and\ \bibinfo {author} {\bibfnamefont {J.}~\bibnamefont
  {Furthmüller}},\ }\bibfield  {title} {\enquote {\bibinfo {title} {Efficiency
  of ab-initio total energy calculations for metals and semiconductors using a
  plane-wave basis set},}\ }\href {\doibase
  https://doi.org/10.1016/0927-0256(96)00008-0} {\bibfield  {journal} {\bibinfo
   {journal} {Computational Materials Science}\ }\textbf {\bibinfo {volume}
  {6}},\ \bibinfo {pages} {15--50} (\bibinfo {year} {1996})}\BibitemShut
  {NoStop}%
\bibitem [{\citenamefont {Kresse}\ and\ \citenamefont
  {Joubert}(1999)}]{Kresse1999}%
  \BibitemOpen
  \bibfield  {author} {\bibinfo {author} {\bibfnamefont {G.}~\bibnamefont
  {Kresse}}\ and\ \bibinfo {author} {\bibfnamefont {D.}~\bibnamefont
  {Joubert}},\ }\bibfield  {title} {\enquote {\bibinfo {title} {From ultrasoft
  pseudopotentials to the projector augmented-wave method},}\ }\href {\doibase
  10.1103/PhysRevB.59.1758} {\bibfield  {journal} {\bibinfo  {journal} {Phys.
  Rev. B}\ }\textbf {\bibinfo {volume} {59}},\ \bibinfo {pages} {1758--1775}
  (\bibinfo {year} {1999})}\BibitemShut {NoStop}%
\bibitem [{\citenamefont {Bl\"ochl}(1994)}]{Bloch1994}%
  \BibitemOpen
  \bibfield  {author} {\bibinfo {author} {\bibfnamefont {P.~E.}\ \bibnamefont
  {Bl\"ochl}},\ }\bibfield  {title} {\enquote {\bibinfo {title} {Projector
  augmented-wave method},}\ }\href {\doibase 10.1103/PhysRevB.50.17953}
  {\bibfield  {journal} {\bibinfo  {journal} {Phys. Rev. B}\ }\textbf {\bibinfo
  {volume} {50}},\ \bibinfo {pages} {17953--17979} (\bibinfo {year}
  {1994})}\BibitemShut {NoStop}%
\bibitem [{\citenamefont {Perdew}, \citenamefont {Burke},\ and\ \citenamefont
  {Ernzerhof}(1996)}]{Perdew1996}%
  \BibitemOpen
  \bibfield  {author} {\bibinfo {author} {\bibfnamefont {J.~P.}\ \bibnamefont
  {Perdew}}, \bibinfo {author} {\bibfnamefont {K.}~\bibnamefont {Burke}}, \
  and\ \bibinfo {author} {\bibfnamefont {M.}~\bibnamefont {Ernzerhof}},\
  }\bibfield  {title} {\enquote {\bibinfo {title} {Generalized gradient
  approximation made simple},}\ }\href {\doibase 10.1103/PhysRevLett.77.3865}
  {\bibfield  {journal} {\bibinfo  {journal} {Phys. Rev. Lett.}\ }\textbf
  {\bibinfo {volume} {77}},\ \bibinfo {pages} {3865--3868} (\bibinfo {year}
  {1996})}\BibitemShut {NoStop}%
\bibitem [{\citenamefont {Stukowski}(2009)}]{OVITO}%
  \BibitemOpen
  \bibfield  {author} {\bibinfo {author} {\bibfnamefont {A.}~\bibnamefont
  {Stukowski}},\ }\bibfield  {title} {\enquote {\bibinfo {title} {Visualization
  and analysis of atomistic simulation data with ovito-the open visualization
  tool},}\ }\href {\doibase 10.1088/0965-0393/18/1/015012} {\bibfield
  {journal} {\bibinfo  {journal} {Modelling and Simulation in Materials Science
  and Engineering}\ }\textbf {\bibinfo {volume} {18}},\ \bibinfo {pages}
  {015012} (\bibinfo {year} {2009})}\BibitemShut {NoStop}%
\bibitem [{\citenamefont {Momma}\ and\ \citenamefont {Izumi}(2011)}]{VESTA}%
  \BibitemOpen
  \bibfield  {author} {\bibinfo {author} {\bibfnamefont {K.}~\bibnamefont
  {Momma}}\ and\ \bibinfo {author} {\bibfnamefont {F.}~\bibnamefont {Izumi}},\
  }\bibfield  {title} {\enquote {\bibinfo {title} {{{\it VESTA3} for
  three-dimensional visualization of crystal, volumetric and morphology
  data}},}\ }\href {\doibase 10.1107/S0021889811038970} {\bibfield  {journal}
  {\bibinfo  {journal} {Journal of Applied Crystallography}\ }\textbf {\bibinfo
  {volume} {44}},\ \bibinfo {pages} {1272--1276} (\bibinfo {year}
  {2011})}\BibitemShut {NoStop}%
\bibitem [{Note1()}]{Note1}%
  \BibitemOpen
  \bibinfo {note} {We note here that with the earlier version of the ML-QUIP
  potential$^{36}$ the order of phases coesite-\MakeUppercase []{ii} and
  coesite-\MakeUppercase []{iii} is correct, the transition pressure being $30$
  GPa, and we were able to find coesite-\MakeUppercase []{iii} by plain MD
  simulation at $35$ GPa and $300$ K.}\BibitemShut {Stop}%
\bibitem [{\citenamefont {Erhard}\ \emph {et~al.}(2022)\citenamefont {Erhard},
  \citenamefont {Rohrer}, \citenamefont {Albe},\ and\ \citenamefont
  {Deringer}}]{Erhard2022}%
  \BibitemOpen
  \bibfield  {author} {\bibinfo {author} {\bibfnamefont {L.~C.}\ \bibnamefont
  {Erhard}}, \bibinfo {author} {\bibfnamefont {J.}~\bibnamefont {Rohrer}},
  \bibinfo {author} {\bibfnamefont {K.}~\bibnamefont {Albe}}, \ and\ \bibinfo
  {author} {\bibfnamefont {V.~L.}\ \bibnamefont {Deringer}},\ }\bibfield
  {title} {\enquote {\bibinfo {title} {A machine-learned interatomic potential
  for silica and its relation to empirical models},}\ }\href {\doibase
  10.1038/s41524-022-00768-w} {\bibfield  {journal} {\bibinfo  {journal} {npj
  Computational Materials}\ }\textbf {\bibinfo {volume} {8}},\ \bibinfo {pages}
  {90} (\bibinfo {year} {2022})}\BibitemShut {NoStop}%
\end{thebibliography}
%

\end{document}